\documentclass[aps,prl,twocolumn,showpacs,superscriptaddress]{revtex4-1}
\usepackage{latexsym}
\usepackage{amssymb}
\usepackage{graphicx}
\usepackage{amsmath}
\usepackage{bm}
\usepackage[colorlinks,
          linkcolor=black,
            citecolor=black,
            urlcolor=blue
           ]{hyperref}
\usepackage{verbatim}
\usepackage{mathrsfs}
\usepackage{extarrows}
\usepackage{comment}
\usepackage{mathtools,slashed}

\allowdisplaybreaks[4]

\usepackage{cancel}

\usepackage{amsthm}

\usepackage[OT1]{fontenc}

\begin{document}

\title{Zero-point theorems in quantum many-body physics}

\author{Yuan Yao}
\email{smartyao@sjtu.edu.cn}
 \affiliation{Institute of Condensed Matter Physics, School of Physics and Astronomy, Shanghai Jiao Tong University, Shanghai 200240, China}

\begin{abstract}
We propose several zero-point type arguments based on the inevitable zero point(s) of a spectral gap within the quantum spin system phase diagrams in various dimensions.
We consider multi-parameter families of Hamiltonian extending the conventional zero-point theorem that includes only one parameter.
Analogously to the zero-point theorem,
we only impose \textit{model-independent} transformation data upon the parameter boundary,
rather than specifying any low-energy dynamics or response.
We further give a series of conjectures,
which generalize our statements in a uniform way.
Our results give powerful and model-independent constraints on the possible relevant operators for critical phenomena in quantum spin models in arbitrary high dimensions.

\end{abstract}


\maketitle
%
%
%
%

\begin{figure}[h] 
\centering 
\includegraphics[width=0.42\textwidth]{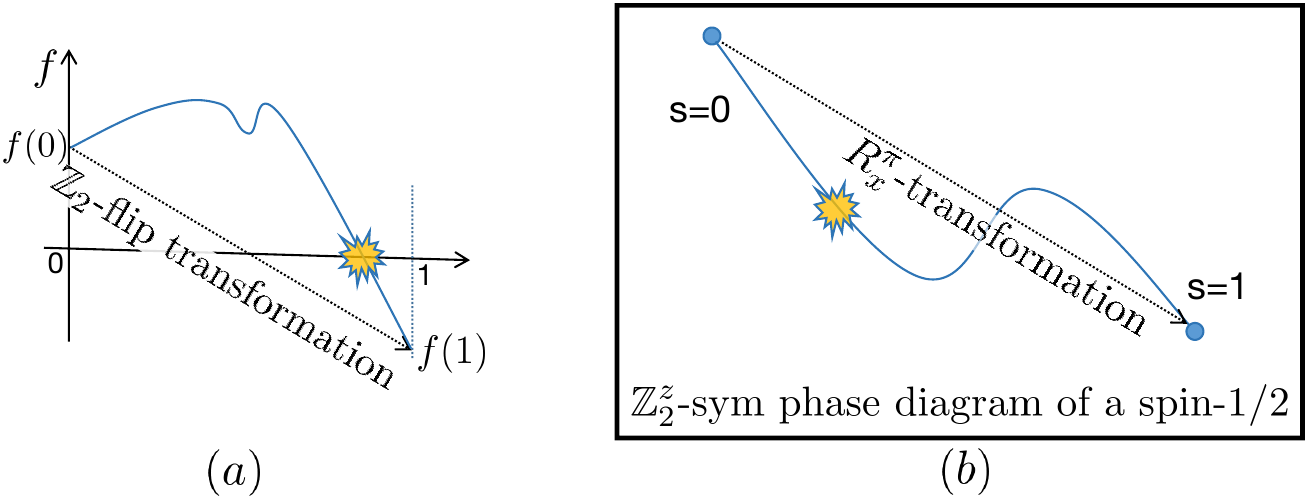} 
\caption{(a) Zero-point theorem: the constraint is a $\mathbb{Z}_2$-transformation on the interval boundary;
(b) Zero-point theorem in quantum mechanics: a gap-closing point (denoted by the burst shape) is inevitable.}\label{zero_thm}
\end{figure}

\paragraph{Introductions and Illustrations.---}
The zero-point theorem states that
\textit{any continuous real function $f$ on the close interval $I=[0,1]$ must possess at-least one zero point as long as $f(1)=-f(0)$.}
Despite its intuitive clearness as in FIG.~\ref{zero_thm}~(a),
it has the following nontrivial aspects:
\begin{itemize}
\item A1: The constraint on the function is only imposed on the interval boundary $\partial I$,
but the bulk behavior is restricted. 
\item A1':
the constraint is purely a \textit{model-independent} symmetry data; it does not specify concrete values of $f$ at $\partial I=\{0,1\}$,
but simply a model-independent $\mathbb{Z}_2$-flip operation.

\item A2: The existence of the zero point(s) is ``topological''; no longer how we deform the function,
it must exist.
\item A3: The position of the zero point(s) is non-universal and,
for a fixed $s\in[0,1]$,
we can always find a function that is nonzero at $s$;
no point in $I$ is compulsory to be a zero point.
\end{itemize}

In quantum physics,
given a many-body Hamiltonian $H$,
there is an important real-valued gap function $\Delta$:
\begin{eqnarray}
\Delta(H)\equiv\lim_{L\rightarrow\infty} E_2(H)-E_1(H),
\end{eqnarray}
where $\lim_{L\rightarrow\infty}$ means the thermodynamic limit and $E_i(H)$ denotes,
in ascending order,
the $i$-th eigenvalue of $H$ at a finite system size $L$.
$\Delta(H)=0$ generically relates to long-range entanglements,
e.g., spontaneously symmetry breaking (SSB),
topological order,
or gapless criticality.
Such zeros or gap closings can also signal the topological quantum phase transition between two short-range entangled phases in which $\Delta\neq0$.
(Let us call the Hamiltonians $H$ with $\Delta(H)\neq0$ as ``$\Delta$-gapped''.)
Thus,
in the quantum phase classifications,
it is interesting to find zero-point-type theorems for the gap function $\Delta$,
which powerfully constrain quantum phase diagrams.

Based on symmetry,
there is a notable theorem given by Lieb, Schultz and Mattis (LSM)~\cite{Lieb:1961aa} and its extensions~\cite{OYA1997,Oshikawa:2000aa,Hastings:2004ab,NachtergaeleSims,Yao:2021aa};
any [$\mathbb{Z}_2\times\mathbb{Z}_2$] spin-rotation symmetric spin chain Hamiltonian $H$ under the periodic boundary condition (PBC)  cannot be $\Delta$-gapped,
i.e., $\Delta(H)=0$,
once $H$ respects a lattice translation with spin-1/2 per unit cell.
The LSM theorem is ``topological'',
but the constraint on the phase diagram is too strong so that any Hamiltonian obeying the above microscopic symmetry data is compulsory to be a zero point of $\Delta$.
This reflects the conventional \textit{pointwise} perspective;
the question we often raise is ``whether a \textit{single} Hamiltonian $H$ satisfies $\Delta(H)=0$ or not''.
The aspect A3 provides a new \textit{family} viewpoint;
``given a family of Hamiltonians,
whether there must exist a Hamiltonian $H$ therein with $\Delta(H)=0$.''
Inspired by such considerations,
we give an illustrating theorem in quantum mechanics as in FIG.~\ref{zero_thm}~(b):

{\bf Theorem~0: }\textit{(Quantum-mechanical zero-point theorem) Given any one-parameter family of single spin-1/2 Hamiltonians $H_s: s\in[0,1]$ respecting $\mathbb{Z}_2^z$ symmetry: $[H_s,R^\pi_z]=0$,
then
\begin{eqnarray}
F: F(s)\equiv\Delta(H_s)
\end{eqnarray}
must possess a zero point if the boundary is related by $\mathbb{Z}_2^x$-transformation:
\begin{eqnarray}
\label{boundary_0}
H_1=R^\pi_x H_0 \left(R^\pi_x\right)^\dagger.
\end{eqnarray}
}
Here $R^\pi_{x,z}=\sigma^{x,z}$ are $\pi$-rotation along $x$- and $z$-axes,
respectively.
The proof is elementary~\footnote{Its proof is elementary.
Due to $\mathbb{Z}_2^z$ symmetry of $H_s$,
the only possibility (up to an identity operator) is $H_s=f(s)\sigma^z$ with a continuous function $g$,
thereby $F(s)=2|f(s)|$.
In addition,
the $\mathbb{Z}_2^x$-relation imposes $f(1)=-f(0)$.
We completes the proof by applying the zero-point theorem on $f$ and the zero point of $f$ and $F$ is the same.},
where Eq.~\eqref{boundary_0} plays exactly the same role as $f(1)=-f(0)$,
and the above theorem instrinsically reflects all the aspects of the zero-point theorem
(although it is a simple single-body problem).

Similar phenomena have been proposed based on response field theory~\cite{Kikuchi:2017aa,Tanizaki:2018aa,Cordova:2020ab,Hsin:2020aa},
where,
however,
a certain concrete form of low-energy field action along the parameter boundary must be specified, 
instead of a model-independent symmetry data as above.
Even the recognization of those low-energy theories on lattices is difficult in general~\cite{Yao:2022vh,Wen:2021aa}.
Such model dependences restrict their application beyond certain constructions,
as we will discuss later in concrete applications.
 
In this \textit{Letter},
we propose several \textit{multi-parameter} zero-point-type theorems in quantum {many-body} systems in various dimensions,
where the interval $I=[0,1]$ is generalized to its higher-dimensional correspondence $I_p\equiv[0,1]^p=[0,1]\times[0,1]\cdots\times[0,1]$ with an integer $p\geq1$.
These statements impose only transformation relation upon the boundary $\partial I_p$,
which extends Eq.~\eqref{boundary_0}.
Thus,
no knowledge of infrared low-energy physics of those Hamiltonians on $\partial I_p$ is pre-required beyond the symmetry data,
as in the aspect A1.
In addition,
there is no LSM-type constraint pointwisely in $I_p$,
so the aspect A3 is respected;
otherwise it is unnecessary to consider a Hamiltonian family.
We also propose several conjectures extending the statements uniformly.
Furthermore,
our results impose powerful constraints on the possible relevant operators of conformal field theories (CFTs) in arbitrary dimensions,
which are largely unclear before.

\paragraph{Warm-up: zero-point theorem in spin-1/2 chains.---}
{\bf Theorem~0} can be understood as an inevitable quantum phase transition as follows.
Assuming $\Delta(H_{0,1})\neq0$
(otherwise the proof is done),
we can conclude that their own unique ground states satisfy
$R^\pi_z|\text{gs}_{0,1}\rangle=\eta_{0,1}|\text{gs}_{0,1}\rangle$ by $R^\pi_z$ symmetry.
In addition,
$|\text{gs}_{0,1}\rangle$ are related as $|\text{gs}_0\rangle=R^\pi_x|\text{gs}_1\rangle$ by Eq.~\eqref{boundary_0},
so $\eta_1=-\eta_0$ by $\sigma^z\sigma^x=(-1)\sigma^x\sigma^z$.
Thus,
the gap of $H_s$ must be closed somewhere $s\in[0,1]$ to host such a quantum phase transition characterizing this $R^\pi_z$-eigenvalue jump.

Such a proof of {\bf Theorem~0} provides a general \textit{proving technique} applicable to many-body physics;
if a Hamiltonian family possesses $G_0$ symmetry ($\mathbb{Z}_2^z$ above),
then we need a transformation $g_1$ ($R^\pi_x$ above) that changes \textit{any} $G_0$-SPT phase to another distinct $G_0$-SPT phase.
It motivates us to propose

%

{\bf Theorem~1: }\textit{(Zero-point theorem in one dimension) Given any one-parameter family of spin-1/2 chains $H_s: s\in[0,1]$ respecting $\mathbb{Z}_2^z\times\mathbb{Z}_2^x$ symmetry,
then
$F: F(s)\equiv\Delta(H_s)$
must possess a zero point in $[0,1]$ if the boundary is related by:
\begin{eqnarray}
\label{boundary_1}
H_1=T H_0 T^\dagger,
\end{eqnarray}
where $\mathbb{Z}_2^{x,z}$ transformations are
generated by $R^\pi_{x,z}=\prod_{j=1}^L\sigma^{x,z}_j$,
and
$T: T\vec{S}_jT^\dagger=\vec{S}_{j+1}$ is the lattice translation.
}


It is sufficient to show that $T$ transformation will change one $[\mathbb{Z}_2^z\times\mathbb{Z}_2^x]$-SPT phase to the other, as follows.

We recall that $[\mathbb{Z}_2^z\times\mathbb{Z}_2^x]$-SPT phases are characterized by the response to the symmetry twisting;
we twist the boundary condition by,
say $R^\pi_x$,
and calculate the $R^\pi_z$-eigenvalue change of the ground state after the twisting.
Then we need to prove such a change is different between $H_0$ and $H_1$.

Specifically,
let us denote the twisted Hamiltonians of $H_{0,1}$ with ground states $|\text{gs}_{0,1}\rangle$ by $H^\text{tw}_{0,1}$ with ground states $|\text{gs}^\text{tw}_{0,1}\rangle$.
Their $R^\pi_z$-eigenvalues are labelled by $\eta_{0,1}$ and $\eta_{0,1}^\text{tw}$.
The boundary twisting is imposed on the link between the $L$-th and $1$-st lattice points,
so the Hamiltonian term across this link will be modified,
e.g., $\vec{S}_{L}\cdot\vec{S}_{1}$ in the original Hamiltonian $H$ becomes $\vec{S}_{L}\cdot \left[R^\pi_x\vec{S}_{1} {R^\pi_x}^\dagger\right]$ in the twisted Hamiltonian $H^\text{tw}$.

Since $H_1=TH_0T^\dagger$,
we have $|\text{gs}_1\rangle=T|\text{gs}_0\rangle$.
Thus,
before the twisting, 
$\eta_1=\eta_0$ because $[R^\pi_z,T]=0$.
However,
$H_1^\text{tw}\neq TH^\text{tw}_0T^\dagger$ due to the non-commutativity between the twisting and $T$;
the twisting in $TH^\text{tw}_0T^\dagger$ is imposed on the link between $1$-st and $2$-nd lattice points while the twisting in $H^\text{tw}_1$ is, by definition, between $L$-th and $1$-st lattice points.
Thus,
\begin{eqnarray}\label{tw_1}
H_1^\text{tw}=\sigma^x_1TH^\text{tw}_0T^\dagger\left(\sigma^x_1\right)^\dagger,
\end{eqnarray}
where $\sigma^x_1$ at 1-st lattice point restores the twisting position back~\cite{Yao:2021aa}.
Then $|\text{gs}^\text{tw}_1\rangle=\sigma^x_1T|\text{gs}^\text{tw}_0\rangle$,
which implies that:
\begin{eqnarray}
\eta^\text{tw}_1&\equiv&\langle\text{gs}^\text{tw}_1|R^\pi_z|\text{gs}^\text{tw}_1\rangle=\langle\text{gs}^\text{tw}_0|T^\dagger\sigma^x_1\left(\prod_j\sigma^z_j\right)\sigma^x_1T|\text{gs}^\text{tw}_0\rangle\nonumber\\
&=&\langle\text{gs}^\text{tw}_0|T^\dagger(-1)\left(\prod_j\sigma^z_j\right)T|\text{gs}^\text{tw}_0\rangle=-\eta_0^\text{tw},
\end{eqnarray}
where the crucial minus sign comes from the anticommutation between $\sigma^x_1$ and $\sigma^z_1$ in $R^\pi_z$.
Combining it with $\eta_0=\eta_1$,
we can compare their $R^\pi_z$-responses:
\begin{eqnarray}
\frac{\eta^\text{tw}_1}{\eta_1}=(-1)\frac{\eta^\text{tw}_0}{\eta_0},
\end{eqnarray}
which means that $H_{0,1}$ must belong to distinct $[\mathbb{Z}_2^z\times\mathbb{Z}_2^x]$-SPT phases.
It completes the proof of {\bf Theorem~1}.

\begin{figure}[t] 
\centering 
\includegraphics[width=0.48\textwidth]{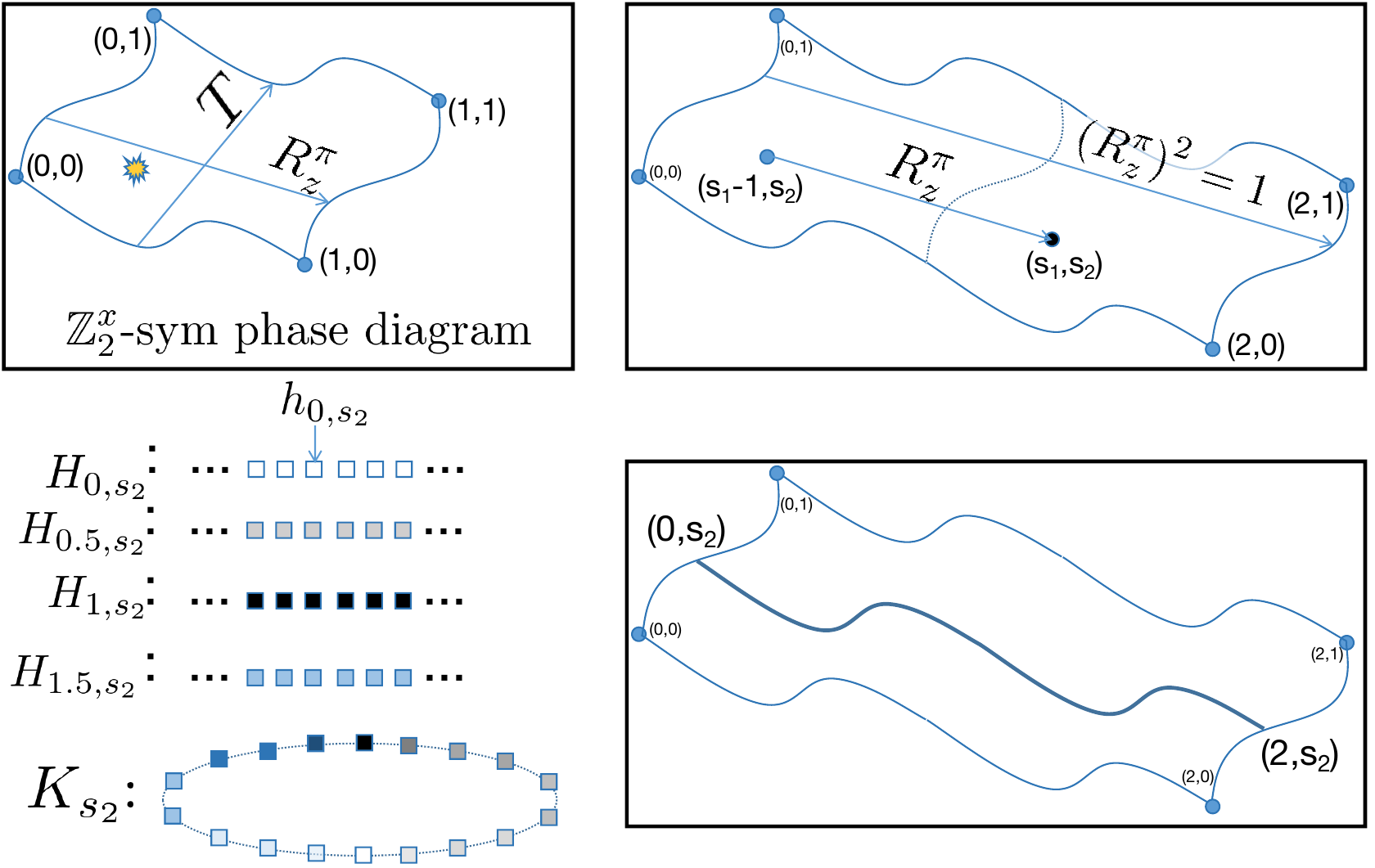} 
\caption{Left-up: {\bf Theorem~2}; Right-up: the doubling procedure  $s_1\in[0,2]$; Bottom: the construction of the texture Hamiltonian $K_{s_2}$.}\label{zero_thm_2}
\end{figure}

{\bf Theorem~1} implies a \textit{model-independent} way to produce a nontrivial $[\mathbb{Z}_2\times\mathbb{Z}_2]$-SPT spin-1/2 Hamiltonian; 
starting from any $[\mathbb{Z}_2\times\mathbb{Z}_2]$-symmetric (e.g., trivial SPT) $H$,
we simply do a $T$-transformation, and either $H$ or $THT^\dagger$ must be nontrivial,
reflecting the power of model-independence aspect~(A1').


\paragraph{Multi-parameter zero-point theorems.---}
The above zero-point theorem can be still understood by the traditional phase transitions,
but inapplicable to multiple-parameter situations as in FIG.~\ref{zero_thm_2}:

{\bf Theorem~2: }\textit{($2$-parameter zero-point theorem in one dimension)
Given any 2-parameter family of spin-1/2 chains $H_{s_1,s_2}: (s_1,s_2)\in[0,1]\times[0,1]$ respecting $\mathbb{Z}_2^x$-symmetry,
then
$F: F(s)\equiv\Delta(H_s)$
must possess a zero point in $[0,1]^2$ if the boundary is related through:
\begin{eqnarray}
\label{boundary_2}
\left\{\begin{array}{l}H_{1,s_2}=R^\pi_z H_{0,s_2} (R^\pi_z)^\dagger\\
H_{s_1,1}=T H_{s_1,0} T^\dagger.
\end{array}\right.
\end{eqnarray}
}
We prove it by contradiction;
we assume the entire family is simultaneously $\Delta$-gapped: $\Delta(H_{s_1,s_2})\neq0, \forall (s_1,s_2)$.

Firstly,
we extend the $s_1$-interval to $[0,2]$ by:
\begin{eqnarray}\label{ext_1}
H_{s_1,s_2}\equiv R^\pi_zH_{s_1-1,s_2}\left(R^\pi_z\right)^\dagger,\,\,(s_1,s_2)\in[0,2]\times[0,1],
\end{eqnarray}
so that $H_{2,s_2}=H_{0,s_2}$ by Eq.~\eqref{boundary_2} and $\left(R^\pi_z\right)^2=1$,
as shown in FIG.~\ref{zero_thm_2}.

We label the local Hamiltonian term of $H_{s_1,s_2}$ by $h_{s_1,s_2}(j)$,
e.g., $h_\text{Ising}(j)=-\sigma_j^z\sigma_{j+1}^z$ for the Ising model. 
Due to the consistency of the chain length $L$ with PBC,
\begin{eqnarray}\label{consis_PBC}
h_{s_1,s_2}(j+L)=T^Lh_{s_1,s_2}(j)T^{-L}.
\end{eqnarray}
In terms of local form,
Eq.~\eqref{boundary_2} under the extension~\eqref{ext_1} is equivalent to
\begin{eqnarray}
\label{boundary_2_local}
\left\{\begin{array}{l}h_{s_1+1,s_2}(j)=R^\pi_z h_{s_1,s_2}(j) (R^\pi_z)^\dagger\\
h_{s_1,1}(j)=Th_{s_1,0}(j-1)T^\dagger.
\end{array}\right.
\end{eqnarray} 
1-family  $H_{s_1,s_2}:s_1\in[0,1]$ with $s_2$ fixed
enables us to construct a texture Hamiltonian $K_{s_2}$ of a chain length $2L$ by local terms as
\begin{eqnarray}\label{local_2}
k_{s_2}(j)\equiv h_{\frac{j-s_2}{L},s_2}(j),\,\,j\in[1,2L],
\end{eqnarray}
under PBC as illustrated in FIG.~\ref{zero_thm_2}.
Intuitively,
$K_{s_2}$ looks like $H_{\frac{j-s_2}{L},s_2}$ when we focus around the $j$-th lattice point.

Varying $s_2$,
we obtain an adiabatic connection between $K_0$ and $K_1$.
Combining the second line of Eq.~\eqref{boundary_2_local} with Eq.~\eqref{local_2},
we have
\begin{eqnarray}
k_{1}(j)=h_{\frac{j-1}{L},1}(j)=Th_{\frac{j-1}{L},0}(j-1)T^\dagger=Tk_0(j-1)T^\dagger,\nonumber
\end{eqnarray}
which gives
\begin{eqnarray}
K_1=TK_0T^\dagger.
\end{eqnarray}
Furthermore,
using Eq.~\eqref{consis_PBC},
the first line of Eq.~\eqref{boundary_2_local} and Eq.~\eqref{local_2},
we obtain
\begin{eqnarray}
k_{s_2}(j+L)&=&T^Lh_{\frac{j-s_2}{L}+1,s_2}(j)T^{-L}\nonumber\\
&=&T^LR^\pi_z\underbrace{h_{\frac{j-s_2}{L},s_2}(j)}_{k_{s_2}(j)}\left(T^LR^\pi_z\right)^\dagger.
\end{eqnarray}
Thus
we find that $K_{s_2}$ respects an enlarged symmetry:
\begin{eqnarray}
G_K=\langle R^\pi_x,T^LR^\pi_z\rangle.
\end{eqnarray}
To achieve a contradiction,
we show that $K_{1,0}$ must belong to distinct $G_K$-SPT phases,
which can be done similarly to {\bf Theorem~1} with the only difference that $R^\pi_z$ symmetry is replaced by $T^LR^\pi_z$.
We use $R^\pi_x$ to twist the boundary link between $2L$-th and $1$-st sites.
The twisted Hamiltonians have the ground states related by $|\text{gs}^\text{tw}_1\rangle=\sigma^x_1T|\text{gs}^\text{tw}_0\rangle$ by similar analyses below Eq.~\eqref{tw_1}.
However,
the symmetry $T^LR^\pi_z$ is no longer preserved by the twisted Hamiltonian $K^\text{tw}_{0,1}$ due to that the non-onsite part $T^L$ moves the twisted link by $L$ sites.
To cancel this movement,
the non-onsite symmetry needs a modification:
\begin{eqnarray}
G_{K^\text{tw}}=\left\langle R^\pi_x,\left(\prod_{j=1}^L\sigma^x_j\right)T^LR^\pi_z\right\rangle.
\end{eqnarray}
Focusing on the $(T^LR^\pi_z)$-responses denoted by $\eta$ again,
we obtain
\begin{eqnarray}
\frac{\eta_1^\text{tw}}{\eta_1}&\equiv&\frac{\langle\text{gs}_1^\text{tw}|\left(\prod_{j=1}^L\sigma^x_j\right)T^LR^\pi_z|\text{gs}_1^\text{tw}\rangle}{\langle\text{gs}_1|T^LR^\pi_z|\text{gs}_1\rangle}=(-1)\frac{\eta^\text{tw}_0}{\eta_0},\nonumber
\end{eqnarray}
where the minus sign comes from $\sigma^x_1$ in the relation $|\text{gs}^\text{tw}_1\rangle=\sigma^x_1T|\text{gs}^\text{tw}_0\rangle$ and $\sigma^z_1$ in $R^\pi_z$ above.
This distinct topological response completes the proof of {\bf Theorem~2}.

\begin{figure}[h] 
\centering 
\includegraphics[width=0.48\textwidth]{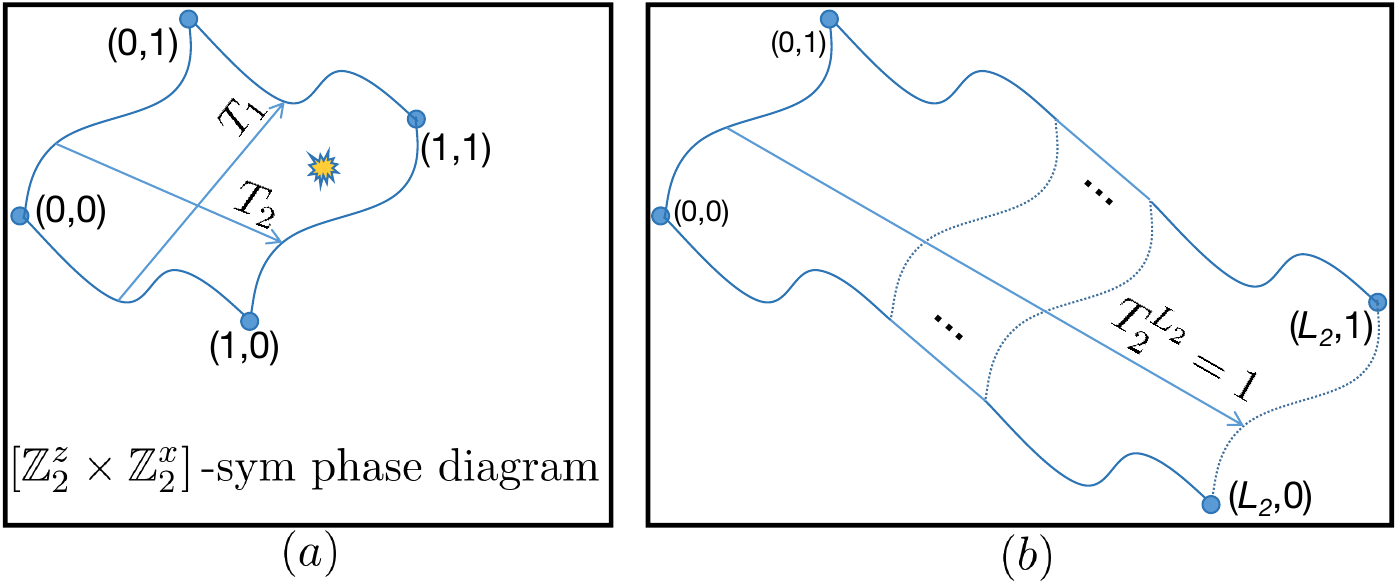} 
\caption{(a) {\bf Theorem~3}; (b) Extension $s_1\in[0,L_2]$.}\label{zero_thm_3}
\end{figure}

Although the above technique makes use of an inevitable phase transition between $K_0$ and $K_1$,
the statement itself is beyond the conventional quantum phase transition between two Hamiltonians in the phase diagram.
We also propose a zero-point theorem in higher dimensions as in FIG.~\ref{zero_thm_3}~(a):

{\bf Theorem~3: }\textit{($2$-parameter zero-point theorem in 2 dimensions)
Given any 2-parameter family of spin-1/2 models on a square lattice $H_{s_1,s_2}: (s_1,s_2)\in[0,1]\times[0,1]$ respecting $\left[\mathbb{Z}_2^z\times\mathbb{Z}_2^x\right]$-symmetry,
then
$F: F(\vec{s})\equiv\Delta(H_{\vec{s}})$
must possess a zero point in $[0,1]^2$ if the boundary is related through:
\begin{eqnarray}
\label{boundary_3}
\left\{\begin{array}{l}H_{1,s_2}=T_2 H_{0,s_2} T_2^\dagger\\
H_{s_1,1}=T_1 H_{s_1,0} T_1^\dagger,
\end{array}\right.
\end{eqnarray}
where $T_{1,2}$ are the two single-lattice translations.
}

\begin{figure}[h] 
\centering 
\includegraphics[width=0.48\textwidth]{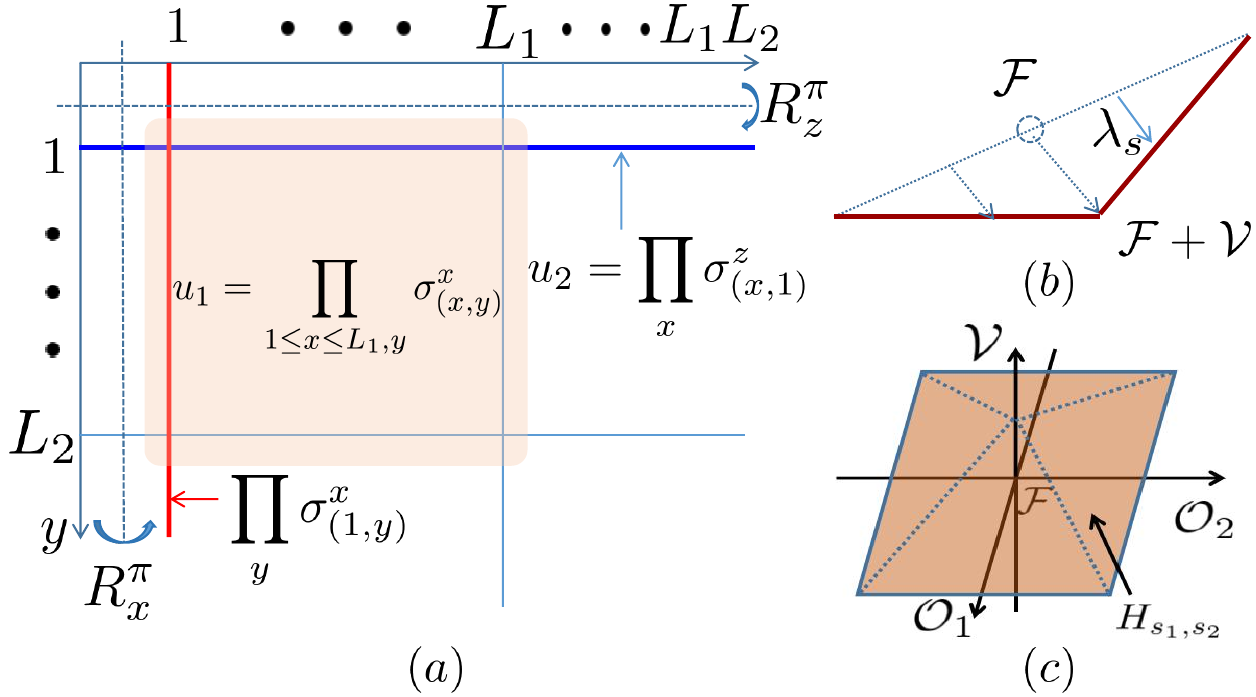} 
\caption{(a) $K^\text{tw}_{s_2}$: The links perpendicularly intersecting the horizontal line $y=1/2$ are twisted by $R^\pi_z$,
and $R^\pi_x$ for those links intersecting the $x=1/2$ line;
$u_2$ intersects $\prod_y\sigma^x_{(1,y)}$ once to contribute $(-1)$ in Eq.~\eqref{anti_3}.
(b) $\Delta$-gapped 1-family ${H}_{s}=\mathcal{F}+2(s-1/2)\mathcal{O}+\lambda_s\mathcal{V}$ with $\lambda_{1/2}=1,\lambda_{0,1}=0$.
(c) $\Delta$-gapped 2-family ${H}_{s}=\mathcal{F}+2(s_2-1/2)\mathcal{O}_1+2(s_1-1/2)\mathcal{O}_2+\lambda_{s_1,s_2}\mathcal{V}$ with $\lambda_{1/2,1/2}=1$ and it vanishes when approaching the boundary.}\label{twist}
\end{figure}

We first extend the $s_1$-interval as in FIG.~\ref{zero_thm_3}~(b)
\begin{eqnarray}
H_{s_1,s_2}=T_2H_{s_1-1,s_2}T_2^\dagger,\,\,(s_1,s_2)\in[0,L_2]\times[0,1],
\end{eqnarray}
so that $H_{L_2,s_2}=H_{0,s_2}$ since $H_{s_1,s_2}$ are put on a $L_1\times L_2$ square lattice under PBC: $T_2^{L_2}=1$.
We construct a texture Hamiltonian $K_{s_2}$ on a $(L_1L_2)\times L_2$ square lattice with local terms:
\begin{eqnarray}
k_{s_2}(x,y)=h_{\frac{x-s_2}{L_1},s_2}(x,y),\,\,\,\,(x,y)\in[1,L_1L_2]\times[1,L_2],\nonumber
\end{eqnarray}
under PBC.
Since,
by the local form of Eq.~\eqref{boundary_3},
\begin{eqnarray}
k_{1}(x,y)&=&h_{\frac{x-1}{L_1},1}(x,y)=T_1h_{\frac{x-1}{L_1},0}(x-1,y)T_1^\dagger\nonumber\\
&=&T_1k_0(x-1,y)T_1^\dagger,
\end{eqnarray}
we obtain
\begin{eqnarray}
K_1=T_1K_0T_1^\dagger.
\end{eqnarray}
Besides the original onsite symmetry $\mathbb{Z}_2^z\times\mathbb{Z}_2^x$,
the symmetry of $K_{s_2}$ is enlarged to
\begin{eqnarray}
G_K=\langle R^\pi_{x,z},T_1^{L_1}T_2\rangle,
\end{eqnarray}
since
\begin{eqnarray}
&&k_{s_2}(x+L_1,y+1)=T^{L_1}h_{\frac{x-s_2}{L_1}+1,s_2}(x,y+1)T^{-L_1}\nonumber\\
&=&\left(T^{L_1}T_2\right)\underbrace{h_{\frac{x-s_2}{L_1},s_2}(x,y)}_{k_{s_2}(x,y)}\left(T^{L_1}T_2\right)^\dagger.
\end{eqnarray}
Let us twist the boundary condition as FIG.~\ref{twist}~(a);
we use $R^\pi_z$ to twist the links perpendicularly intersecting the horizontal line $y=1/2$,
and $R^\pi_x$ for those links intersecting the $x=1/2$ line.
The symmetry group is modified to
\begin{eqnarray}
G_{K^\text{tw}}=\left\langle R^\pi_{x,z},(u_1T_1^{L_1})(u_2T_2)\right\rangle,
\end{eqnarray}
where $u_{1,2}$ are to compensate the non-onsite effects from $T_1^L$ and $T_2$, respectively: $
u_1\equiv\prod_{y,1\leq x\leq L_1}\sigma^x_{(x,y)};\,\,u_2\equiv\prod_{x}\sigma^z_{(x,1)}$ as shown in FIG.~\ref{twist}.
The twisted ground states of $K_{0,1}^\text{tw}$ are related by $|\text{gs}^\text{tw}_1\rangle=\prod_y\sigma^x_{(1,y)}T_1|\text{gs}_0^\text{tw}\rangle$.
We study their $(T_1^LT_2)$-responses denoted by $\eta$:
\begin{eqnarray}\label{anti_3}
&&\frac{\eta_1^\text{tw}}{\eta_1}\equiv\frac{\langle\text{gs}_1^\text{tw}|u_1T^{L_1} _1u_2T_2|\text{gs}_1^\text{tw}\rangle}{\langle\text{gs}_1|T_1^{L_1} T_2|\text{gs}_1\rangle}=(-1)\frac{\eta_0^\text{tw}}{\eta_0},
\end{eqnarray}
where $u_1'\equiv T_1^\dagger u_1T_1=\prod_{y,0\leq x\leq L_1-1}\sigma^x_{(x,y)}$ and the essential $(-1)$ comes from the anticommutation between the intersection between $\prod_y\sigma^x_{(1,y)}$ in the relation $|\text{gs}^\text{tw}_1\rangle=\prod_y\sigma^x_{(1,y)}T_1|\text{gs}_0^\text{tw}\rangle$ and $u_2$ as in FIG.~\ref{twist}~(a).
Therefore,
we reach a contradiction with the $G_K$-symmetric adiabatic connection between $K_0$ and $K_1$.
It completes the proof of {\bf Theorem~3}.

\paragraph{Generalizing conjectures and revisiting the LSM theorem.---}
We propose general statements to summarize and envelop all the theorems so far in a uniform way.

{\bf Conjecture~4: }\textit{Given any $p$-parameter family of spin-1/2 models on a $d$-dimensional hyper-cubic lattice $H_{\vec{s}}: \vec{s}\in[0,1]^p$ with $p\leq d$ respecting $\left[\mathbb{Z}_2^z\times\mathbb{Z}_2^x\times T_{p+1}\times\cdots\times  T_d\right]$-symmetry,
then
$F: F(\vec{s})\equiv\Delta(H_s)$
must possess a zero point in $[0,1]^p$ if the boundary is related through $(j=1,\cdots,p)$:
\begin{eqnarray}
\label{boundary_d}
H_{s_1,\cdots,1_j,\cdots,s_p}=T_j H_{s_1,\cdots,0_j,\cdots,s_p} T_j^\dagger,
\end{eqnarray}
where $T_j$ is the lattice translations along $j$-th direction.
}

We also conjecture a generalization with one $\mathbb{Z}_2$ of the $\mathbb{Z}_2^z\times\mathbb{Z}_2^x$ moved to the boundary transformation:

{\bf Conjecture~5: }\textit{
Given any $(d+1)$-parameter family of spin-1/2 models on a $d$-dimensional hyper-cubic lattice $H_{\vec{s}}: \vec{s}\in[0,1]^p$ with $p\leq d+1$ respecting $\left[\mathbb{Z}_2^x\times T_p\times\cdots\times T_d\right]$-symmetry,
then
$F: F(\vec{s})\equiv\Delta(H_s)$
must possess a zero point in $\vec{s}\in[0,1]^p$ if the boundary is related through $(j=1,\cdots,p-1)$:
\begin{eqnarray}
\label{boundary_d}
\left\{\begin{array}{l}H_{s_1,\cdots,1_j,\cdots,s_p}=T_j H_{s_1,\cdots,0_j,\cdots,s_p} T_j^\dagger,\\
H_{s_1,s_2,\cdots,1_{p}}=R^\pi_z H_{s_1,s_2\cdots,0_p} \left(R^\pi_z\right)^\dagger.\end{array}\right.
\end{eqnarray}
}

We justify these conjectures by using them to reproduce the LSM theorem in higher dimensions.
Given a $d$-dimensional $\left[\mathbb{Z}_2^z\times\mathbb{Z}_2^x\times T_1\times\cdots\times T_d\right]$-symmetric spin-1/2 system $H$,
then the constant family $H_{\vec{s}\in[0,1]^d}\equiv H$ trivially satisfies Eq.~\eqref{boundary_d},
thereby $\Delta[H]=0$ by {\bf Conjecture~4}.
Thus,
our results should contain many more useful constraints than the LSM theorem as below.

\paragraph{Constraints on conformal field theories and discussions.---}
For a CFT with Hamiltonian density $\mathcal{F}$,
if a perturbation $\mathcal{O}$ gaps the system so that $\Delta[\mathcal{F}+\mathcal{O}]>0$,
we call it $\Delta$-gapping.
Let us see how these renormalization-group fixed points are restricted.
\begin{itemize}
\item
If $\mathcal{F}$ is realized by a
spin-$1/2$ chain respecting $\left[\mathbb{Z}_2^z\times\mathbb{Z}_2^x\times T\right]$ and $\mathcal{F}$ has one $[\mathbb{Z}^z_2\times\mathbb{Z}^x_2]$-symmetric $\Delta$-gapping relevant $T$-odd direction:
$T\mathcal{O}T^\dagger=-\mathcal{O}$,
then we claim that it cannot have any other $[\mathbb{Z}^z_2\times\mathbb{Z}^x_2]$-symmetric  (including marginally) $\Delta$-gapping relevant operator $\mathcal{V}$.
If there were such $\mathcal{V}$,
we consider a 1-family $H_{s}=\mathcal{F}+2(s-1/2)\mathcal{O}+\lambda_{s}\mathcal{V}$,
where the real function $\lambda_{1/2}=1$ and $\lambda_{0}=\lambda_{1}=0$ as in FIG.~\ref{twist}~(b).
Then $H_{s}$ satisfies the condition of {\bf Theorem~2},
but it is $\Delta$-gapped for all $s$,
contradicting with {\bf Theorem~2}.

\item
If $\mathcal{F}$ is realized by a 2-dimensional
spin-$1/2$ square lattice respecting $\left[\mathbb{Z}_2^z\times\mathbb{Z}_2^x\times T_1\times T_2\right]$ and has $[\mathbb{Z}^z_2\times\mathbb{Z}^x_2]$-symmetric $\Delta$-gapping relevant directions $\mathcal{O}_{1,2}$,
which are $T_{1,2}$-odd, respectively:
$\mathcal{O}_jT_j=-T_j\mathcal{O}_j$,
then we can
similarly conclude that it cannot have any other $[\mathbb{Z}^z_2\times\mathbb{Z}^x_2]$-symmetric  (including marginally) $\Delta$-gapping relevant direction $\mathcal{V}$ by FIG.~\ref{twist}~(c) with its contradiction to {\bf Theorem~3}.
\item 
{\bf Conjectures~4,5} similarly constrains the CFTs in arbitrary higher dimensions,
which are largely unclear before.
\end{itemize}

Let us see concrete examples for the one-dimensional case above.
$[\mathbb{Z}_2\times\mathbb{Z}_2\times T]$-symmetric critical spin-1/2 chains realize odd-level SU(2)$_{2l+1}$ Wess-Zumino-Witten CFTs~\cite{Affleck:1987aa,Affleck:1988aa,Furuya:2017aa,Yao:2019aa}with the fundamental matrix field $g$, and $TgT^\dagger=-g$.
The most relevant $[\mathbb{Z}_2\times\mathbb{Z}_2]$-symmetric operator $\mathcal{O}_{1/2,1/2}=$Tr$(g)$,
which is also $T$-odd,
is expected to be $\Delta$-gapping to a valence-bond-state phase.
By our constraint,
the marginally relevant current-current operator $J_\text{L}\cdot J_\text{R}$ must gap the system to a (e.g., resonant-valence-bond) SSB phase~\cite{Affleck:1987aa},
and any other primary field $\mathcal{V}_{J,J}:J\in[1,(2l+1)/2]\cap\mathbb{Z}/2$,
after projected to $[\mathbb{Z}_2\times\mathbb{Z}_2]$-singlet,
must be either irrelevant,
or gapping the CFT to some SSB phase or flowing to another CFT.
It provides many more constraints than LSM theorem that only restricts integral $J$ since $T\mathcal{V}_{J,J}T^\dagger=(-1)^{2J}\mathcal{V}_{J,J}$.

The \textit{model-independence} should re-emphasized that,
although the \textit{proof} of {\bf Theorem~2} uses the idea of inevitable phase transitions,
we do not require the CFTs above are phase-transition points between distinct SPT phases,
in contrast to earlier model-dependent results~\cite{Hsin:2020aa,Yao:2022vh} that can only give nontrivial constraints for specified CFTs. 
Model-dependence discussion can be similarly made for {\bf Theorem~3} and {\bf Conjectures~4,5}.


\paragraph{Acknowledgements.---}
The author thank Chenjie Wang for inspiring and constructive advices and Xinliang Lyu for the suggestion of the title. I am also grateful to Shang-Qiang Ning, Masaki Oshikawa,
Xiao-Qi Sun,
and Rong Yu for helpful discussions.
	The work of Y.~Y. was supported by the National Key Research and Development Program of China (Grant No.~2024YFA1408303),
	the National Natural Science Foundation of China (Grant No.~12474157),
	the sponsorship from Yangyang Development Fund,
	and Xiaomi Young Scholars Program. 


%

\end{document}